\documentclass[letter,dvipsnames]{aa}
\usepackage[utf8]{inputenc}
\usepackage[T1]{fontenc}
\usepackage{txfonts}
\usepackage{graphicx}
\usepackage{natbib}
\bibpunct{(}{)}{;}{a}{}{,} % to follow the A&A style

\begin{document}
\titlerunning{Li-rich giant in NGC\,1261}
\title{The Gaia-ESO Survey: an extremely Li-rich giant\\in the globular cluster NGC\,1261\thanks{Based on data obtained with the ESO VLT under the observing programmes 188.B-3002, 193.B-0936, and 197.B-1074.}}
\author{N.~Sanna\inst{\ref{oaa}}, 
        E.~Franciosini\inst{\ref{oaa}}, 
        E.~Pancino\inst{\ref{oaa},\ref{ssdc}},
        A.~Mucciarelli\inst{\ref{unibo},\ref{oabo}},
        M.~Tsantaki\inst{\ref{oaa}},
        C.~Charbonnel\inst{\ref{cc1},\ref{cc2}},
        R.~Smiljanic\inst{\ref{rs}},
        X.~Fu\inst{\ref{xf}},
        A.~Bragaglia\inst{\ref{oabo}},
        N.~Lagarde\inst{\ref{utinam}},
        G.~Tautvai\v siene\inst{\ref{lt}},
        L.~Magrini\inst{\ref{oaa}},
        S.~Randich\inst{\ref{oaa}},
        T.~Bensby\inst{\ref{lund}},
        A.~J.~Korn\inst{\ref{uppsala}},
        A.~Bayo\inst{\ref{IFA},\ref{NPF}},
        M.~Bergemann\inst{\ref{MPIA}},
        G.~Carraro\inst{\ref{unipd}},
        L.~Morbidelli\inst{\ref{oaa}}
        }

%NOTA: usa le labels per gli istituti cosi' non ci si confonde quando ci sono troppi autori
\institute{INAF - Osservatorio Astrofisico di Arcetri, Largo E. Fermi 5, 50125 Firenze, Italy\label{oaa}
\and ASI Science Data Center, Via del Politecnico SNC, I-00133 Rome, Italy\label{ssdc}
\and Dipartimento di Fisica e Astronomia, Universit\`a degli Studi di Bologna, via Gobetti 93/2, I-40129 Bologna, Italy\label{unibo}
\and INAF - Osservatorio di Astrofisica e Scienza dello Spazio di Bologna, via Gobetti 93/3, I-40129 Bologna, Italy \label{oabo}
\and Department of Astronomy, University of Geneva, Chemin des Maillettes 51, 1290 Versoix, Switzerland \label{cc1}
\and
IRAP, UMR 5277, CNRS and Université de Toulouse, 14 Av. E. Belin, 31400 Toulouse, France \label{cc2}
\and Nicolaus Copernicus Astronomical Center, Polish Academy of Sciences, ul. Bartycka 18, 00-716, Warsaw, Poland \label{rs}
\and The Kavli Institute for Astronomy and Astrophysics at Peking University, Beijing
100871, China\label{xf}
\and Institut UTINAM, CNRS UMR 6213, Univ. Bourgogne Franche-Comté, OSU THETA Franche-Comté-Bourgogne, Observatoire
de Besançon, BP 1615, 25010 Besançon Cedex, France\label{utinam}
\and Astronomical Observatory, Institute of Theoretical Physics and Astronomy, Vilnius University, Sauletekio av. 3, 10257 Vilnius, Lithuania \label{lt}
\and Lund Observatory, Department of Astronomy and Theoretical Physics, Box 43, SE-221 00 Lund, Sweden \label{lund}
\and Observational Astrophysics, Division of Astronomy and Space Physics, Department of Physics and Astronomy, Uppsala University, Box 516, SE-751 20 Uppsala, Sweden \label{uppsala}
\and Instituto  de  F\'isica  y Astronom\'ia,  Facultad de Ciencias,  Universidad de Valpara\'iso, Av. Gran Breta\~na 1111, 5030 Casilla, Valpara\'iso, Chile \label{IFA}
\and N\'ucleo Milenio de Formaci\'on Planetaria - NPF, Universidad de Valpara\'iso, Av. Gran Breta\~na 1111, Valpara\'iso, Chile \label{NPF}  
\and Max-Planck Institut f\"{u}r Astronomie, K\"{o}nigstuhl 17, 69117 Heidelberg, Germany \label{MPIA}
\and Dipartimento di Fisica e Astronomia, Universit\`a di Padova, Vicolo dell'Osservatorio 3, 35122 Padova, Italy \label{unipd}}
\authorrunning{N. Sanna et al.}   
\date{Received: September 2019}

\abstract{Lithium rich stars in globular clusters are rare. In fact, only 14 have been found so far, in different evolutionary phases from dwarfs to giants. Different mechanisms have been proposed to explain this enhancement, but it is still an open problem.
Using spectra collected within the Gaia-ESO Survey, obtained with the GIRAFFE spectrograph at the ESO Very Large Telescope,
we present the discovery of the first Li-rich star in the cluster NGC~1261, the second star known in the red giant branch bump phase. The star shows an extreme Li overabundance of A(Li)$_{LTE}=3.92\pm0.14$, corresponding to A(Li)$_{NLTE}=3.40$~dex.
We propose that the Li enhancement is caused by fresh Li production through 
an extra mixing process (sometimes referred to as {\em cool bottom burning}) or could be a pre-existing Li overabundance resulting from binary mass transfer, likely from a red giant branch star, because of the low barium abundance. To unambiguously explain the Li enhancement in globular cluster stars, however, a reliable determination of the abundance of key species like Be, $^6$Li, $^{12}$C/$^{13}$C, and several s-process elements is required, as well as detailed modeling of chromospheric activity indicators.}


\maketitle

%%%%%%%%%%%%%%%%%%%%%%%%%%%%%%%%%%%%%%%%%%%%%%%%%%%%%%%%%%%%%%%%%%%%%%%%%%%%%%%%%%%%%%

\section{Introduction}
\label{sec:intro}

Lithium (Li) is a fragile element: it is synthesised during the Big Bang nucleosynthesis and destroyed at relatively low temperatures 
($\sim2.5\times 10^6$~K) in the stellar interiors \citep{reeves74}. However, classical stellar evolution theory predicts that it is preserved in the stellar envelopes of metal-poor dwarf stars, with subsequent dilution mixing processes during the post-MS evolution.
The investigation of Li abundances in Globular Clusters \citep[GCs, e. g.,][and reference therein] {mucciarelli18} shows a complex situation: most dwarf stars share the same lithium abundance, $A(\mathrm{Li})\simeq2.2$ dex \footnote{A(Li) = log [N(Li)/N(H)] + 12} as found in most halo stars by \citet{spitea, spiteb}, forming the so called Spite plateau. 
Mixing processes during the sub-giant-branch (SGB)
cause a sharp drop in its abundance by about 1~dex \citep[e.g. see Figure~3 in][]{mucciarelli12}. Then, during the red giant branch (RGB), stars between the 
first dredge-up
and the RGB bump form another plateau at about $A(\mathrm{Li})\simeq1.0$ dex \citep{gratton00,mucciarelli12}. After the bump, Li is diluted again \citep{charbonnel2007}. These results have been obtained analysing stars in different evolutionary phases, from the main sequence (MS) up to the asymptotic giant branch (AGB), in only a few GCs with different methods and number of stars. This observational behavior is complicated in the case of GCs by the presence of multiple populations  \citep{bastian18,gratton19}. In fact, in some of the proposed scenarios the polluting material is almost Li-free and once diluted with Li-normal gas should produce a lower Li abundance, but this is only found in some clusters \citep[e.g., NGC\,6752,][]{pasquini6752} and not in others, \citep[e.g., NGC\,362,][]{dorazi362all}, where first and second population stars share the same Li abundance, or in $\omega$ Cen, where some second population stars have the same Li abundance as the first population and others do not \citep[see][and reference therein, for details]{mucciarelli18}.

%------------------------
\begin{figure}[t]
  \centering
  \resizebox{\hsize}{!}{\includegraphics[clip]{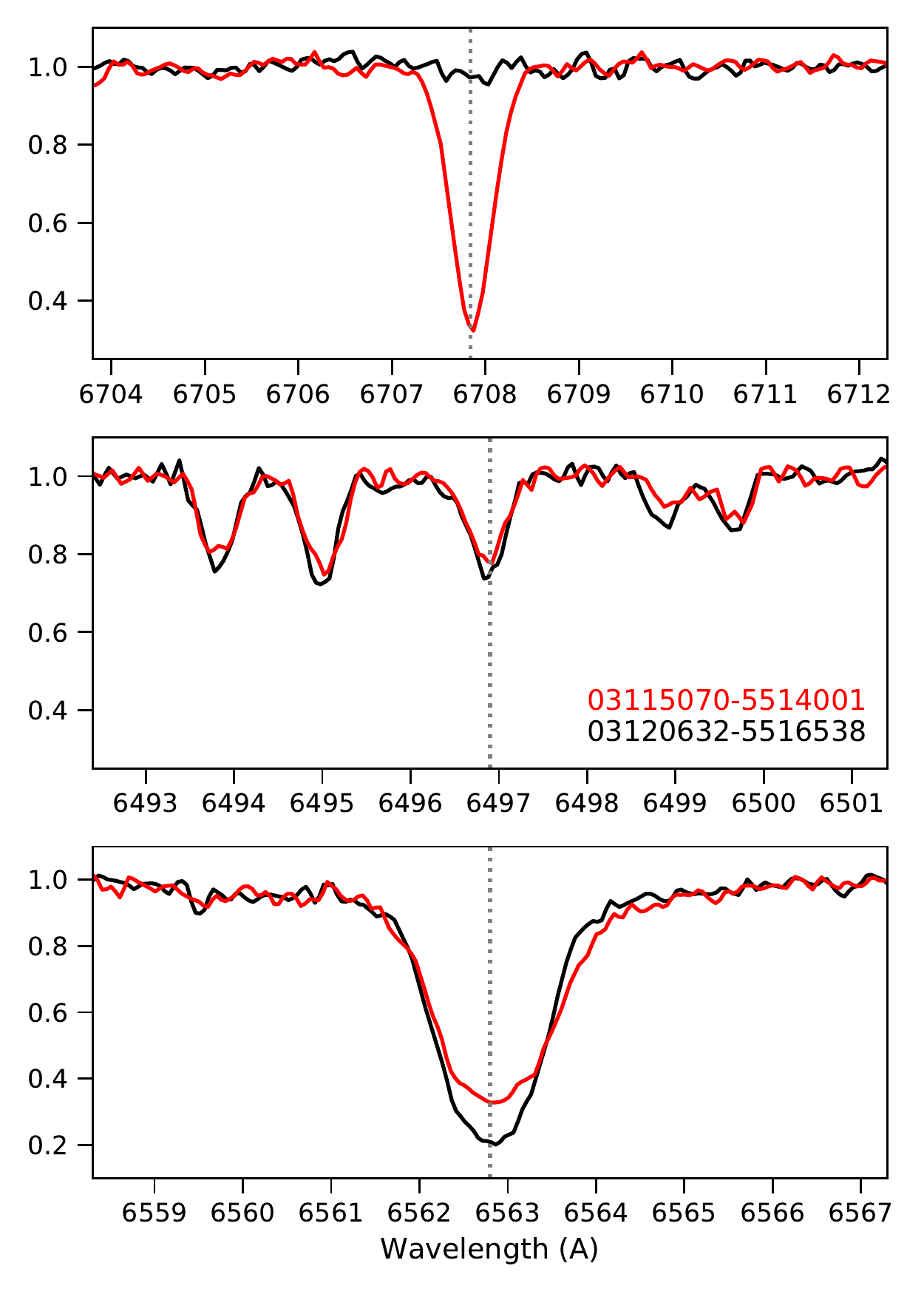}}
  \caption{Comparison of the spectrum of the Li-rich giant in NGC\,1261 (red) with
  that of a normal giant with similar parameters (black), in the regions containing the Li doublet at 6707.8\AA\ (top panel), the Ba\,II line at 6497\AA\ (middle panel) and the H$_{\alpha}$ line (bottom panel). The GES IDs of the two stars are indicated in the middle panel. See also Table~\ref{tab:dati}.}
  \label{fig:spectra}
\end{figure}
%-------------------------

To this already quite complex framework, we have to add the existence of 14 Li-rich stars, i.e., with a significantly increased Li abundance ($\gtrsim$0.5~dex) compared to normal GC stars in the same evolutionary phase. These stars are extremely rare \citep[particularly in GCs, but also in open clusters and the field, see i. e.,][]{casey19} and for the moment they have proven to be also extremely difficult to explain. Four main possible enrichment paths have been explored and originally proposed to explain their high Li abundance:
\begin{itemize}
\item{engulfment of substellar systems such as planets or brown dwarfs \citep{siess99}, valid especially for stars with metallicity higher than $\simeq\,$--0.5~dex. In fact several studies \citep[e.g.,][] {gonzalez,fischer,casey16} show that this scenario appears unlikely for metal-poor stars; other studies \citep[e.g.,][]{charbonnel2000} suggest that this scenario is unlikely because $^6$Li, $^7$Li, and Be should be enhanced by the accretion at the same time, contrary to some observations \citep{balachandran00}; moreover the high Li overabundance (A(Li)$\gtrsim$2.2~dex) observed in some stars is difficult to achieve by planet engulfment only \citep{aguilera16}; finally most cases of Li-rich stars should be observed above the RGB bump, where the stars have larger radii and thus planet engulfment is more likely;}
\item{self-enrichment through the Cameron-Fowler mechanism \citep{cameronfowler71}. This could occur in two different evolutionary phases: {\em (i)} extra mixing during the RGB bump phase \citep[sometimes called ``cool bottom processing"][]{boothroyd95,boothroyd99,palacios01} when, thanks to the first dredge-up, the $^3$He-rich envelope has been transported close to the hotter region of the H-burning shell and {\em(ii)} fresh Li production during the AGB, early-AGB or AGB thermal pulses phases, depending on the mass of the star \citep{venturadantona11,charbonnel2000}};
\item{mass-transfer from different kinds of binary companions: {\em (i)} an AGB or super-AGB star, that implies also enrichment in the s-process elements \citep{busso99,venturadantona11}, depending on the AGB star mass or {\em (ii)} an RGB star that produces fresh Li thanks to extra mixing process, but in this case no enrichment in s-process elements should be found;}
\item{ejecta during a nova outburst of a companion star. Numerical simulations predict that a high amount of $^7$Li can be produced when a thermonuclear runaway occurs in the hydrogen envelope of the accreting white dwarf \citep{starrfield}, and this possibility has been confirmed by the discovery of the first $^7$Li detection in a nova \citep{izzo15}. Also a high enrichment of $^7$Be is expected during a nova outburst, as seems confirmed by the first detections \citep[see e. g.][]{molaro}.}
\end{itemize}

%-------------------------------------- 
\begin{table}
\caption{Estimates of relevant parameters for the Li-rich star in NGC\,1261 and a comparison star (see Sect.~\ref{sec:res} for details).}
\label{tab:dati} 
\centering                         
\begin{tabular}{lrr}
\hline\hline
& Star & Star \\
CNAME & 03115070-5514001&03120632-5516538\\
\hline
R.A. (hh mm ss)&03:11:50.70&03:12:06.32\\
Dec (dd mm ss)&--55:14:00.1&--55:16:53.8 \\
SNR & 21 & 27\\
RV (km~s$^{-1}$)& 69.1 $\pm$ 0.2& 69.7 $\pm$ 0.2\\
T$_{\rm{eff}}$ (K)&4904 $\pm$ 77&4835 $\pm$ 76\\
log\,$g$ (dex) &2.3 $\pm$ 0.1 &2.2 $\pm$ 0.1\\
$v_t$ (km~s$^{-1}$) & 1.5$\pm$0.2 & 1.5$\pm$0.2\\
$[$Fe/H$]$ (dex) & --1.27 & --1.27 \\
EW(Li) (m\AA)&350$\pm$25 &20$\pm$5\\
A(Li)$_{\rm{LTE}}$ (dex) &3.92 $\pm$ 0.14 &0.86 $\pm$0.10\\
$\Delta$Li$_{\rm{NLTE}}$ (dex) &--0.52 &+0.11\\
A (Ba II) (dex)&0.69$\pm$0.16&0.96$\pm$0.17\\
\hline\hline
\end{tabular}
%\end{small}
\end{table}
%-------------------------------------- 

In this framework, we present the discovery of a new Li-rich star in the GC NGC\,1261 with Gaia-ESO Survey data. 
NGC\,1261 is a GC in the Horologium constellation, with [Fe/H]=--1.27 dex and low extinction, E(B--V)=0.01~mag \citep{harris10}. It is also one of the few GCs for which an extended
stellar halo was detected \citep{raso20}.
This star is significantly more Li-rich than the bulk of the other Li-rich stars discovered in GCs, similarly to the cases of NGC\,6397 \citep{koch11}, NGC\,4590 \citep[M\,68]{kraft99}, and NGC\,5272 \citep[M\,3,][]{ruchti11}. 

%%%%%%%%%%%%%%%%%%%%%%%%%%%%%%%%%%%%%%%%%%%%%%%%%%%%%%%%%%%%%%%%%%%%%%%%%%%%%%%%%%%%%%

\section{Data analysis and results}
\label{sec:res}

We based our work on the Gaia-ESO survey \citep[hereafter GES,][]{gilmore,randich} spectra. In particular, we used the FLAMES-GIRAFFE \citep{pasquini} spectra obtained with the HR15N (647--679~nm) setup at medium resolution (R=$\lambda/\delta\lambda\simeq$20\,000), which includes the Li doublet at 6707.8\AA. The GES spectra reductions for GIRAFFE were performed with in-house software as described in detail by \citet{jackson15}, resulting in one single spectrum for each star, combining different observations. The combined spectrum of the discovered Li-rich star in NGC\,1261, with GES CNAME 03115070-5514001, is plotted in Fig.~\ref{fig:spectra} along with a comparison star in the same cluster and with similar properties, 03120632-5516538. The huge difference in the strength of the Li line at 6707.8\AA\ is evident. The Li-rich star and the comparison star have a radial velocity of about 69.1 and 69.7~km~s$^{-1}$, respectively, fully compatible with the systemic velocity of NGC\,1261 \citep[68.2 $\pm$ 4.6~km~s$^{-1}$,][]{harris10}.

%------------------------
\begin{figure}[t]
    \centering
    {\includegraphics[width=0.9\columnwidth]{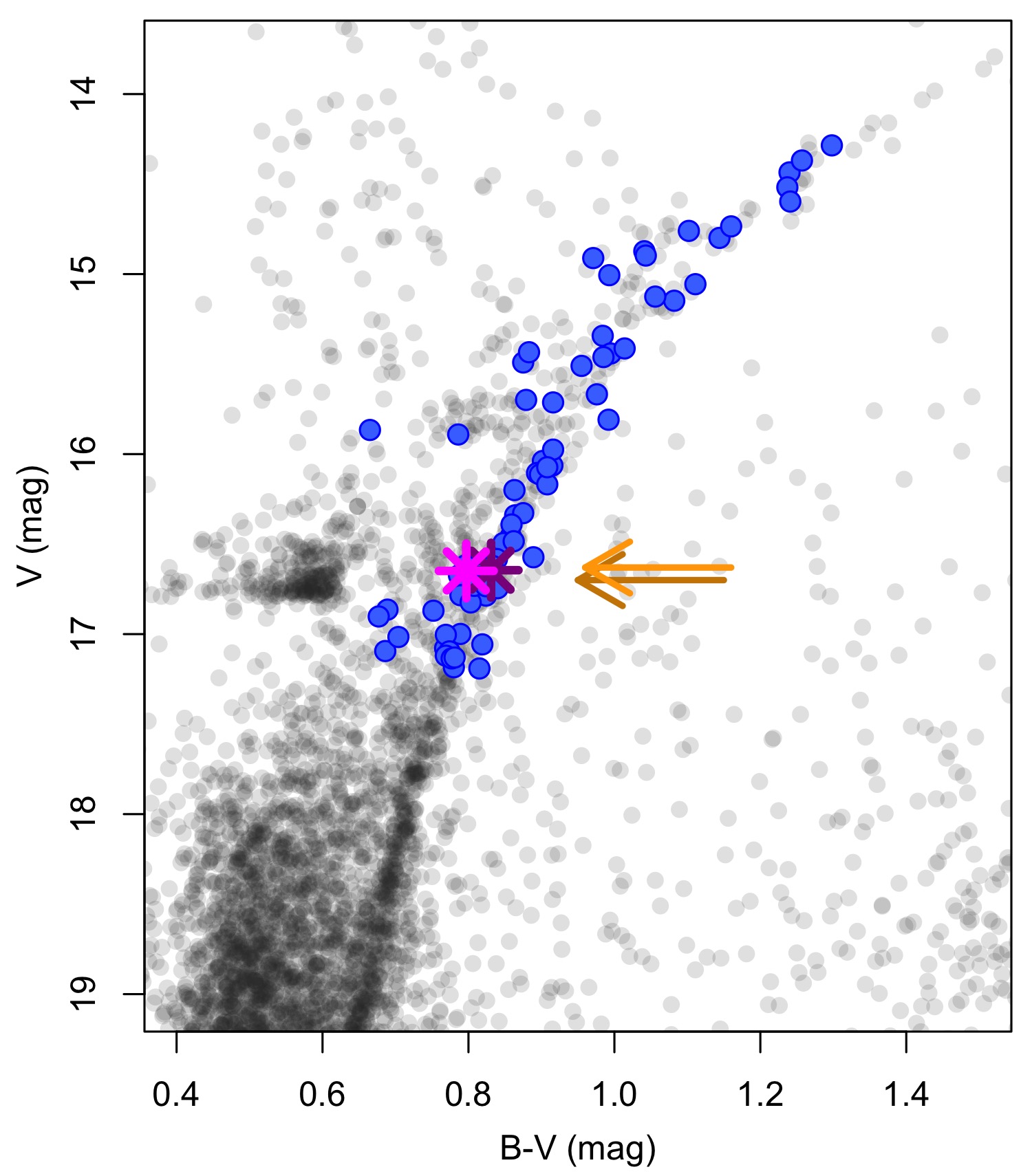}}
    \caption{The CMD of NGC\,1261 from \citet[grey points]{stetson19} with the position of the Li-rich star (magenta star) and the comparison star (purple star). The other stars observed by GES with GIRAFFE are plotted in blue. The position of the RGB bump is also indicated, as determined by \citet[][brown arrow]{ferraro93} and \citet[][orange arrow]{kravtsov10}.}
    \label{fig:cmds}
\end{figure}
%-------------------------

The two stars were observed in November 2017, therefore they are not present in the publicly available GES DR3 release\footnote{\url{https://www.eso.org/qi/}}, based on data observed until July 2014, nor in the
current internal release, iDR5, which is based on spectra acquired until December 2015. For the purpose of the present work, we have preliminarly estimated the relevant quantities (Table~\ref{tab:dati}). The equivalent width (EW) of the Li 6707.8\AA\ doublet was measured manually with IRAF\footnote{IRAF is the Image Reduction and Analysis Facility, a general purpose software system for the reduction and analysis of scientific data. IRAF is written and supported by the IRAF programming group at the National Optical Astronomy Observatories (NOAO) in Tucson, Arizona.}. As it can be seen, the star displays a Li absorption line that is more than 100 times \textbf{stronger} than that of a typical RGB star in NGC\,1261. The \citet{stetson19} photometry was used to evaluate the star's position in the color-magnitude diagram (CMD), showing that it lies precisely on the RGB-bump (Fig.~\ref{fig:cmds}), using the bump estimates by \citet{ferraro93} and \citet{kravtsov10}. Using B--V and V--I colors from the \citet{stetson19} photometry and the color-temperature calibrations by \citet{alonso99,alonso01}, we estimated the effective temperature T$_{\rm{eff}}$ and the surface gravity log\,$g$ (Table~\ref{tab:dati}).

We computed the Li abundance for the two stars using the spectral synthesis code SALVADOR developed by one of the authors (AM), based on the Kurucz abundance analysis routines \citep{kur93a,kur93b,sbordone04} and using 1D ATLAS9 models, which assume local thermodynamical equilibrium (LTE). We assumed the \citet{harris10} metallicity for NGC\,1261 of [Fe/H]=--1.27~dex and a fixed micro-turbulent velocity of $v_t=$1.5~km~s$^{-1}$, derived using the latest version of the Gaia-ESO calibration \citep[][Worley et al., in prep.]{smiljanic14}. We varied the $^6$Li and $^7$Li abundance in the region of the Li doublet at 6707.8\AA\  to search for the best fit. For the comparison star, changing $v_t$ or the isotopic ratio implied variations of $\lesssim$0.01~dex in the final Lithium abundance. For the Li-rich star, a 0.3~km~s$^{-1}$ variation of $v_t$ implied a variation of 0.03~dex on the final Li abundance, and it was included in the final uncertainty computation as well as the uncertainties on T$_{\rm{eff}}$ and log\,$g$. Most importantly, as observed by others, the choice of a solar $^6$Li/Li$_{\rm{tot}}$ ratio of 0.075 caused a poor fit of the asymmetric line profile and a final Li abundance 0.42~dex lower. In the case of Trumpler\,5, a $^6$Li/Li$_{\rm{tot}}\lesssim$2\% was found also when using 3D modeling \citep{monaco14}. Assuming no $^6$Li, the fit of the line profile was significantly improved: in this case the final Li abundance is A(Li)$_{\rm{LTE}}$=3.92$\pm$0.14~dex, almost 1000 times higher than the typical RGB star at the level of the bump. We also report in Table~\ref{tab:dati} the non-LTE correction to the lithium abundance for the two stars, obtained using the data by \citet{lind09a}\footnote{\url{http://inspect.coolstars19.com/}}.

The HR15N setup includes also the barium (Ba~II) line at 6497~\AA\ and the H$_{\alpha}$ Balmer line at 6563~\AA\ (middle and bottom panels of Fig.~\ref{fig:spectra}, respectively). Using the same spectral synthesis setup used for Li, we obtained A(Ba)~$=0.69 \pm 0.16$ dex and A(Ba)~$=0.96 \pm 0.17$ dex for the Li-rich star and the comparison star, respectively. As can be seen from the bottom panel of Fig.~\ref{fig:spectra}, the H$_{\alpha}$ line of the Li-rich star is significantly shallower than the reference star and shows a marked asymmetry on the red wing, that is quite different than in normal RGB stars. This anomalous profile is compatible with chromospheric activity and mass loss \citep{meszaros09}. Unfortunately, no asteroseismologic data are available for this star. It would be interesting to measure the projected rotational velocity as well, but with the GIRAFFE resolution we can only observe that there is no significant difference in the profile width of lines between the two stars. Unfortunately the HR15N setup that was analyzed does not contain tracers of multiple populations in GCs. From the (U--B)--(B--I) photometric colour index the star seems to belong to the first population.  

%%%%%%%%%%%%%%%%%%%%%%%%%%%%%%%%%%%%%%%%%%%%%%%%%%%%%%%%%%%%%%%%%%%%%%%%%%%%%%%%%%%%%

%------------------------
\begin{figure}[t]
\centering
\includegraphics[width=\columnwidth]{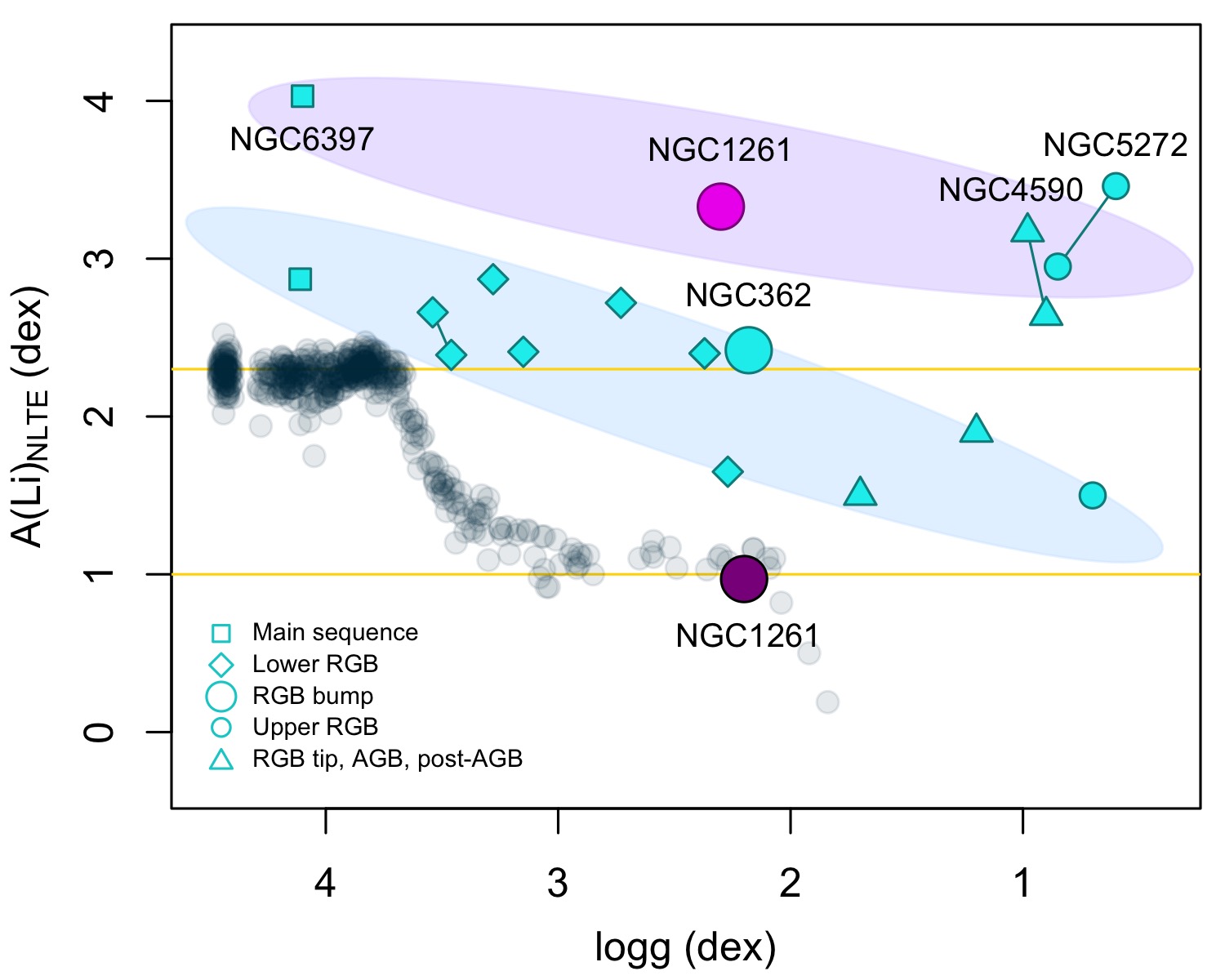}
\caption{Known Li-rich stars in GCs. Literature measurements are plotted as cyan symbols, our measurements for NGC\,1261 are in magenta (for the Li-rich star) and purple (for the comparison star). Evolutionary phases are represented by different shapes as detailed in the legend. Different measurements of the same star are connected by segments. The Li-normal stars in NGC\,6397 by \citet{lind09b} are shown in grey for comparison. The Spite and the low RGB plateaus are shown as yellow lines. Shaded regions represent the main group of Li-rich stars, enhanced by about 0.5--1.5~dex compared to normal stars (cyan region) and the extremely Li-rich stars, about 1.5~dex above them (magenta region).}
\label{fig:lit}
\end{figure}
%-------------------------

\section{Discussion and conclusions}
\label{sec:discussion}

There are 14 Li-rich stars known in GCs, accounting for a few percent of the stars studied spectroscopically: two dwarfs \citep{koch11,monaco12}, two AGB stars \citep{kirby16}; one post-AGB \citep{carney98} and 9 stars along different regions of the RGB or SGB \citep{kraft99,smith99,ruchti11,dorazi15,kirby16, gruyter16, mucciarelli19}. As can be seen from Fig.~\ref{fig:lit}, most of the known Li-rich stars display typically an enhancement of the lithium abundance of 0.5--1.5~dex compared to normal stars in GCs, at least up to the RGB bump. 
Three of the literature stars, however, are significantly more Li-enhanced than the bulk of the sample and than the Spite plateau, by $\Delta$A(Li)$\simeq$1--2~dex, as labelled in Fig.~\ref{fig:lit}. One of these ``super-Li-rich'' stars is the dwarf star in NGC\,6397, discovered by \citet{koch11}, for which two possible origins have been proposed: mass transfer from an RGB companion or engulfment of a sub-stellar system. So far, the star in NGC\,6397 is the only Li-rich star in GCs for which planet engulfment was considered \citep[but see][]{pasquini14}. The other two are bright giants well above the RGB bump, in NGC\,4590 \citep[M\,68,][]{ruchti11,kirby16} and in NGC\,5272 \citep[M\,3,][]{kraft99,ruchti11}. Both the extra mixing process and mass-transfer from a companion have been proposed for these stars.

The Li-rich giant in NGC\,1261 presented here is the fourth super-Li-rich star to be discovered. It clearly belongs to the RGB bump (Fig.~\ref{fig:cmds}), like the Li-rich star in NGC\,362 \citep[][]{dorazi15}, a GC with similar metallicity and horizontal-branch morphology, but the star in NGC\,1261 displays a higher Li abundance: A(Li)$_{\rm{NLTE}}$=3.40$\pm$0.14 as opposed to 2.42$\pm$0.09~dex. The Li abundance of the star in NGC~362 was explained as likely due to the extra mixing process, even if the authors also discuss the possibility that the star could belong to the pre-zero age horizontal branch, implying that it has already experienced the helium flash, because it lies slightly on the blue side or the RGB. This might apply also to our Li-rich star, but both stars appear closer to the RGB than to the horizontal branch. If we consider the extra mixing process, the different A(Li) enrichment of the NGC\,362 and NGC\,1261 stars could be explained by observing that the NGC\,362 star lies just above the RGB bump, where mixing mechanisms could have already started to occur, lowering Li and the $^{12}$C/$^{13}$C ratio. If this is correct, then our Li-rich star in NGC\,1261 should have a higher $^{12}$C/$^{13}$C ratio, i.e., it should have just reached the maximum Li enrichment just before mixing effects extend deep enough to lower the carbon ratio and to start destroying Li \citep{charbonnel2000}.

To understand the origin of the Li-rich star in NGC\,1261, we can count on three pieces of evidence: (1) the profile fit of the Li doublet suggests a very low $^6$Li/Li$_{\rm{tot}}$ ratio; (2) the Ba~II abundance appears low and compatible with that of normal stars in NGC\,1261; and (3) the H$_{\alpha}$ line shows signs of chromospheric activity, i.e., central re-emission and asymmetric profile. The low Ba II abundance seems to suggest that we can exclude mass-transfer from an AGB companion, but we need more s-process elements to confirm that. Moreover, the absence of $^6$Li in our best fit suggests that there has been fresh Li production. With the data in hand, our preferred hypothesis is that the Li enhancement of the RGB bump star in NGC~1261 is made by the extra mixing process, because the star lies precisely on the RGB bump, where this process is expected to occur. Unfortunately, this process does not explain the chromospheric activity suggested by the anomalous H$_{\alpha}$ profile. The engulfment hypothesis of a sub-stellar body seems unlikely because it cannot explain the very high Li enrichment observed in the NGC\,1261 star \citep{siess99}, but it would explain the chromospheric activity. Alternatively, mass transfer from an RGB companion (undergoing the extra mixing process) when the star was on the main sequence could also be a viable hypothesis. If this was the case, similarly to the AGB companion case, then the initial Li abundance of the star -- before the first dredge-up -- should have been of about $\simeq$ 4.6~dex
\footnote{A(Li) $\simeq$ 1~dex for a typical RGB bump star, while here A(Li) $\simeq$ 3.4~dex. 3.4 - 1.0 = 2.4. Adding this value to the Spite plateau (2.2~dex), the star should have an initial abundance $\simeq$ 4.6~dex.}, higher than the Li-rich star in NGC\,6397. 
Moreover, as was recently proposed, other kind of interactions in binary systems could allow the Cameron-Fowler mechanism \citep{casey19}.

Additional information is clearly needed to unambiguously determine the origin of the Li enhancement in our NGC\,1261 star and in the other 14 known Li-rich stars in GCs. In particular, accurate and precise Be and $^6$Li abundances would allow us to evaluate the hypothesis of engulfment of a sub-stellar system. On the other hand, the $^{12}$C/$^{13}$C ratio would be important to distinguish between production of fresh Li through extra mixing process and a pre-existing Li overabundance, perhaps resulting from binary mass transfer \citep{charbonnel2000}. A full determination of s-process abundances could ultimately exclude the mass transfer hypothesis from an AGB companion, or help pinpointing the mass range of the AGB donor \citep{busso99}. Additionally, measurement of the light elements that are known to anti-correlate in GCs (C, N, O, Na, Al, Mg) would shed light on the relation between Li-rich stars and the presence of multiple stellar populations in GCs.
%%%%%%%%%%%%%%%%%%%%%%%%%%%%%%%%%%%%%%%%%%%%%%%%%%%%%%%%%%%%%%%%%%%%%%%%%%%%%%%%%%%%%%

\begin{acknowledgements}
We thank the referee for the useful suggestions. NS and EP acknowledge the financial support to this research by INAF, through the Mainstream Grant 1.05.01.86.22 assigned to the project “Chemo-dynamics of globular clusters: the Gaia revolution” (P.I. E. Pancino). TB was partly funded by the grant 2018-04857 from the Swedish Research Council, and partly by the project grant ’The New Milky Way’ from the Knut and Alice Wallenberg Foundation. Based on data products from observations made with ESO Telescopes at the La Silla Paranal Observatory under programmes ID 188.B-3002, 193.B-0936, and 197.B-1074. These data products have been processed by the Cambridge Astronomy Survey Unit (CASU) at the Institute of Astronomy, University of Cambridge, and by the FLAMES/UVES reduction team at INAF/Osservatorio Astrofisico di Arcetri. These data have been obtained from the Gaia-ESO Survey Data Archive, prepared and hosted by the Wide Field Astronomy Unit, Institute for Astronomy, University of Edinburgh, which is funded by the UK Science and Technology Facilities Council. This work was partly supported by the European Union FP7 programme through ERC grant number 320360 and by the Leverhulme Trust through grant RPG-2012-541. We acknowledge the support from INAF and Ministero dell' Istruzione, dell' Universit\`a' e della Ricerca (MIUR) in the form of the grants "Premiale VLT 2012" and "Premiale 2015 MITIC". The results presented here benefit from discussions held during the Gaia-ESO workshops and conferences supported by the ESF (European Science Foundation) through the GREAT Research Network Programme.

\end{acknowledgements}

%%%%%%%%%%%%%%%%%%%%%%%%%%%%%%%%%%%%%%%%%%%%%%%%%%%%%%%%%%%%%%%%%%%%%%%%%%%%%%%%%%%%%%

\bibliographystyle{aa} % style aa.bst
\bibliography{nsanna} % your references Yourfile.bib

\end{document}